\documentclass[onecolumn,showpacs,preprintnumbers,amsmath,amssymb]{revtex4}

\usepackage{graphicx}
\usepackage{epsfig}

\def\dd{\displaystyle}

\begin{document}
\title{\bf Radiative Correction to the Casimir Energy for Massive Scalar Field on a Spherical Surface}

\author{M. A. Valuyan}
\email{m-valuyan@sbu.ac.ir}
\affiliation{Department of Physics, Semnan Branch, Islamic Azad University, Semnan, Iran}
\date{\today}

\begin{abstract}
In this paper, the first order radiative correction to the Casimir energy for a massive scalar field in the $\phi^4$ theory on a spherical surface with $S^2$ topology was calculated. In common methods for calculating the radiative correction to the Casimir energy, the counter-terms related to free theory are used. However, in this study, by using a systematic perturbation expansion, the obtained counter-terms in renormalization program were automatically position-dependent. We maintained that this dependency was permitted, reflecting the effects of the boundary conditions imposed or background space in the problem. Additionally, along with the renormalization program, a supplementary regularization technique that we named Box Subtraction Scheme (BSS) was performed. This scheme presents a useful method for the regularization of divergences, providing a situation that the infinities would be removed spontaneously without any ambiguity. Analysis of the necessary limits of the obtained results for the Casimir energy of the massive and massless scalar field confirmed the appropriate and reasonable consistency of the answers.
\end{abstract}

\maketitle

\section{Introduction}
\label{sec:intro}
Casimir energy is usually obtained from the difference between the vacuum energy in the presence and absence of non-trivial boundary conditions or non-trivial backgrounds. More than 60 years ago, this effect was predicted and calculated by Hendrik B. Casimir. For the first time in his famous essay, Casimir explained the attraction between two uncharged perfectly conducting parallel plates due to the polarization of electromagnetic field\,\cite{h.b.g.}.
The first attempts to measure this phenomenon and find a credible witness in the laboratory for this effect, nearly ten years later, in 1958, was conducted by M. J. Sparnaay\,\cite{Sparnaay.M.J.} and later, more precise measurements confirmed the correctness of the Casimir prediction\,\cite{Experimental.accurate.1,Experimental.accurate.1.5,Experimental.accurate.2}. Given that the Casimir effect is known as an important effect in reflecting the properties of boundary conditions in quantum field theory, many applications in several areas of physics are reported for it\,\cite{see.some.review.1,see.some.review.2,see.some.review.4,see.some.review.4.5,see.some.review.5,see.some.review.5.5}.
\par
In addition, calculation of radiative correction to the Casimir energy for multiple fields has been considered by physicists. Bordag et al.\,\cite{Bordag.et.al.1,Bordag.et.al.2,bordag.linding.} originally conducted the first order radiative correction known as \emph{two loop correction} to the electromagnetic field. Later, radiative correction to the Casimir energy was conducted for other quantum fields with multiple boundary conditions\,\cite{radiative.correction.free.counterterms.1,radiative.correction.free.counterterms.2,radiative.correction.free.counterterms.3, radiative.correction.free.counterterms.4,radiative.correction.free.counterterms.5}. But in most previous works, the counter-term related to the free theory was used in renormalization program. We maintain that for a problem in which, a non-trivial boundary condition (or non-trivial background) influences the quantum field, performing the renormalization program for that field should be implemented with respect to the new dominant conditions. In fact, it is expected that the translational symmetry breaking should be reflected in the $n$-point function in renormalized perturbation theory. This statement necessities using the position-dependent counter-terms instead of free counter-terms in the renormalization program. This viewpoint on renormalization program has been explained in detail in previous studies\,\cite{posision.dependent.counterterms.works.ours.1,posision.dependent.counterterms.works.ours.1.25,
posision.dependent.counterterms.works.ours.1.5,posision.dependent.counterterms.works.ours.2,posision.dependent.counterterms.works.others.1,posision.dependent.counterterms.works.others.2.5,posision.dependent.counterterms.works.others.2.75,posision.dependent.counterterms.works.others.3} and all their physical aspects and advantages have been discussed. The same idea in the renormalization of interacting quantum field theory in the curved space time has been extensively investigated \,\cite{Birrel and Davis.1,Birrel and Davis.2,renormalization.curved.space.2,renormalization.curved.space.3,renormalization.curved.space.3.5,renormalization.curved.space.4,renormalization.curved.space.5,banach.}. Although, the renormalization counter-terms have been systematically computed  from the components of the energy-momentum tensor in curved space time using several methods\,\cite{hathrell.,bordag.linding.,Renormalization.Group.1,Renormalization.Group.2,Renormalization.Group.3,feynmanrule.1,feynmanrule.2,Toms.}. In this paper, by focusing on the correctness of that idea and using the position-dependent counter-terms in the renormalization program, radiative correction to the Casimir energy for a massive scalar field was studied on a curved manifold. In our calculation procedure, the counter-terms are the same as those in flat case, becoming position-dependent due to the dominant boundary condition on scalar field caused by non-trivial topology of the background space.
\par
One of the most important parts of the calculation of the Casimir energy is regularizing and removing the divergent expressions in appear due to the vacuum energy. Therefore, to achieve this end and regularize and remove these divergences, many methods have been proposed by physicists in this area\,\cite{regularization.methods.1.,regularization.methods.2.,regularization.methods.3.,regularization.methods.3.5,regularization.methods.4.,regularization.methods.5.}. The Box Subtraction Scheme (BSS) as a regularization technique was originally used by T. H. Boyer\,\cite{boyer.} in the calculation of the Casimir energy for electromagnetic field confined in a perfectly conducting sphere. Later, in order to reduce the ambiguities in the removal of divergences, this method was remarkably used in other studies\,\cite{BSS.1,BSS.2,BSS.3}. This technique was also performed to calculate the first order radiative correction to the Casimir energy between two parallel plates for massive scalar field in $\phi^4$ theory in one, two and three spatial dimensions\,\cite{posision.dependent.counterterms.works.ours.1,posision.dependent.counterterms.works.ours.2}. In this paper, the BSS is firstly used for the Casimir energy on a curved manifold. Due to the definition of the Casimir energy, the contribution of the vacuum energy of Minkowski space should be subtracted from the vacuum energy of the desired configuration.  However, the Casimir energy in the BSS is resulted from the subtracting of vacuum energies of two similar configurations in proper limits. These two similar configurations help the appearing infinities to be simply regularized so that its removal process is conducted with clarity.  In the present work, at the first step, using the BSS, the leading order of the Casimir energy density for real massive scalar field in $\phi^4$ theory on a surface with $S^2$ topology is calculated and results are consistent with those reported in previous works\,\cite{new.developements.1,new.developements.2}. In Section 3, using the above-mentioned renormalization program along with the BSS, first order radiative correction to the Casimir energy for $\phi^4$ theory on a surface with $S^2$ topology is also dealt with. Finally, appropriate limits of the obtained answers are discussed and then Section 4 summarizes all physical aspects of using the methods.
\begin{figure}
     \hspace{0cm}\includegraphics[width=8cm]{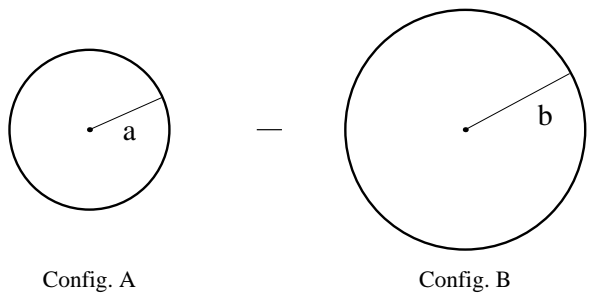}
 \caption{\label{figs.sphere}   Left figure shows a sphere with radius $a$ named \emph{configuration  A}. The right figure shows a similar configuration with a different radius $b$ named \emph{configuration B}. To calculate the Casimir energy, the zero-point energies of these two spherical configurations should be subtracted according to the Eq.\,\eqref{ECasDefinition.}. In the final step, the radius of configuration B goes to infinity, while the other parameters of problem are maintained fixed.}
\end{figure}

\section{Leading Order of Casimir Energy}
\label{sec:Cas.Cal.}
The lagrangian of real massive scalar field with self-interaction term $\phi^4$ is written as follows\,\cite{Birrel and Davis.1,Birrel and Davis.2}:
\begin{equation}\label{lagrangy.}
  \mathcal{L}= \frac{1}{2}\sqrt{-g}\bigg[g^{\mu\nu}\partial_\mu \phi \partial_\nu \phi-(m_{0}^2+\xi \mathcal{R}) \phi^2-\frac{\lambda_0}{4!} \phi^4 \bigg]
\end{equation}
where $m_0$ and $\lambda_0$ are bare mass of the field and bare value of coupling constant, respectively. Furthermore, $\mathcal{R}$ is the scalar curvature of space time and $\xi$ is conformal coupling constant. Moreover, by considering the two-dimensional surface with $S^2$ topology with radius $a$, the scalar curvature $\mathcal{R}=2a^{-2}$ and constant $\xi=\frac{1}{8}$  are obtained and the metric $g^{\mu\nu}$  for this surface is written as follows\,\cite{Birrel and Davis.1,Birrel and Davis.2,new.developements.1,new.developements.2,general relativity}:
\begin{equation}\label{metric.}
  ds^2=dt^2-a^2 (d\theta^2+\sin^2 \theta d\varphi^2 )
\end{equation}
where $x=(t,\theta,\varphi)$ shows the space-time coordinate. To calculate the Casimir energy based on the BSS, two similar configurations should be introduced. Thus, two spheres with radii $a$ and $b$ are introduced. As Fig.\,(\ref{figs.sphere}) shows, we have named these two spheres  \emph{configuration A} and \emph{configuration B}, respectively. Then, vacuum energies of these two configurations are subtracted from each other and in order to find the Casimir energy value for configuration A, the radius of the sphere B should eventually go to infinity. Obviously, to calculate the leading order (zero order of $\lambda_0$) Casimir energy density, the self-interaction term of the lagrangian in Eq.\,(\ref{lagrangy.}) is avoided. Therefore, definition of this order of Casimir energy density based on the BSS can be written as:
\begin{equation}\label{ECasDefinition.}
\mathcal{E}^{(0)}_{\text{Cas.}}=\lim_{b/a\rightarrow\infty}\big[\mathcal{E}^{(0)}_A-\mathcal{E}^{(0)}_B\big]
\end{equation}
where $\mathcal{E}_A^{(0)}$  and $\mathcal{E}_B^{(0)}$ are the vacuum energy density of two similar configurations A and B and the superscript $(0)$ denotes the zero (or leading) order of this energy. In common definitions of the Casimir energy, the contribution of vacuum energy of the Minkowski space is subtracted from vacuum energy of the given configuration in the problem. However, in the BSS, the contribution of the Minkowski space is replaced with a similar configuration (\emph{e.g.} configuration B) that has the properties of the Minkowski space in proper limit ($b\rightarrow\infty$). In fact, selecting  two configurations in the BSS caused the useful parameters to be imported in the calculation procedures. These parameters usually play the role of  a regulator, causing the divergences to be removed clearly and possible ambiguities in the calculation process to be reduced.
\par
After canonical quantization of the field and using the equation of motion extracted from the lagrangian shown in Eq.\,(\ref{lagrangy.}), vacuum energy density on spherical surface will be obtained for the real scalar field with periodic boundary condition. This energy density for the configuration A is as follows\,\cite{new.developements.1,new.developements.2}:
\begin{equation}
\label{vacuum.energy}
\mathcal{E}^{(0)}_A=<0\mid T_{00}\mid0>=\frac{1}{4\pi a^2}\sum_{\ell=0}^{\infty}\bigg(\ell+\frac{1}{2}\bigg)\omega_{\ell}
\end{equation}
where $\omega_\ell=\left[\frac{1}{a^2}\left(\ell+\frac{1}{2}\right)^2+m_{0}^2\right]^{\frac{1}{2}}$ is the wave number. Similarly, by replacing radius $a$ with $b$ in the above equation, vacuum energy density for configuration B is obtained. To calculate the Casimir energy from Eq.\,(\ref{ECasDefinition.}), the difference between the two divergent expressions $\mathcal{E}_{A}^{(0)}$ and $\mathcal{E}_B^{(0)}$ should be computed. Therefore, Abel-Plana Summation Formula (APSF) for half integer parameters is used\,\cite{Generalized.Abel.Plana.Saharian}. This formula converts all the summation expressions into the integral form and for the expression within the brackets in Eq.\,(\ref{ECasDefinition.}), after using the APSF, we have:
\begin{eqnarray}\label{Vacuum.En.Diff.}
   \mathcal{E}^{(0)}_A-\mathcal{E}^{(0)}_B=\frac{1}{4\pi a^2}\bigg\{\frac{1}{a}\int_{0}^{\infty}z[m_{0}^2a^2+z^2]^{\frac{1}{2}}dz+B(a)\bigg\}-\{a\rightarrow b\}
\end{eqnarray}
The second term in square brackets is named \emph{Branch-cut} term and its value is: $\dd B(x)=2m_{0}^3 x^2 \int_0^1 \frac{z\sqrt{1-z^2}}{e^{2\pi m x z}+1}dz$ . The integration result for the Branch-cut term is finite, while the first term in both square brackets in Eq.\,\eqref{Vacuum.En.Diff.} is divergent. In order to remove infinities due to these two terms, at the first step, the upper limit of these integrations in Eq.\,\eqref{Vacuum.En.Diff.} is substituted for $\Lambda_a$ and $\Lambda_b$. Then, by calculating integrations, we will have an answer as a function of cutoffs $\Lambda_a$ and $\Lambda_b$, respectively. It can be shown that, by selecting proper values for $\Lambda_a$ and $\Lambda_b$ in subtraction process in Eq.\,\eqref{Vacuum.En.Diff.}, all of the infinities are canceled and the only remaining terms from Eq.\,\eqref{Vacuum.En.Diff.} are Branch-cut terms. Therefore, we have,
\begin{eqnarray}\label{Branch-cut.Diff}
   \mathcal{E}^{(0)}_A-\mathcal{E}^{(0)}_B=\frac{m_{0}^3}{2\pi}\int_0^1 \frac{z\sqrt{1-z^2}}{e^{2\pi m_{0} a z}+1}dz-\{a\rightarrow b\}
\end{eqnarray}
Finding a closed form for the result of the above integral is extremely difficult or might be impossible. Therefore, by expanding the denominator of integrals in Eq.\,\eqref{Branch-cut.Diff} and calculating them, we have,
\begin{eqnarray}\label{Casimir:Energy}
 \mathcal{E}^{(0)}_A-\mathcal{E}^{(0)}_B=\frac{m_{0}^3}{2\pi}\sum_{j=1}^{\infty}(-1)^{j+1}\int_0^1 z\sqrt{1-z^2}e^{-2\pi m_0 a z j}dz-\{a\rightarrow b\}
  \nonumber\hspace{4cm} \\=\sum_{j=1}^{\infty}\frac{(-1)^{j+1}\mu^2}{24\pi j a^3}\left[4\mu j-3 I_2\left(2\pi \mu j\right)+3 L_2\left(2\pi \mu j\right)\right]-\{a\rightarrow b\}\hspace{3.1cm}
\end{eqnarray}
where $\mu=m_{0}a$ and functions $I_2(x)$ and $L_2(x)$ are modified Bessel function and modified struve function, respectively. At the last step, the limit stated in Eq.\,\eqref{ECasDefinition.} is calculated and the leading order to the Casimir energy density for massive scalar field on a spherical surface with radius $a$ is obtained:
\begin{equation}\label{Leading.order.CasEn.}
  \mathcal{E}^{(0)}_{\textrm{Cas.}}=\sum_{j=1}^{\infty}\frac{(-1)^{j+1}\mu^2}{8\pi j a^3}\bigg[\frac{4\pi^2 j^2 \mu^2-3}{3\pi^2 j \mu}-I_2\left(2\pi \mu j\right)+L_2\left(2\pi \mu j\right)\bigg]
\end{equation}
As shown in Fig.\,(\ref{figs.Casimir.}), this energy has good consistency with the expected physical grounds so that it is in agreement with what exists in Refs.\,\cite{new.developements.1,new.developements.2}. It should be noted that the obtained result have been automatically reached by the BSS as a regularization technique and fortunately there is no need to perform any renormalization program manually (See for instance Section (3.4) of Ref.\,\cite{new.developements.1,new.developements.2}).

\section{First Order Radiative Correction}
\par
\begin{figure}
     \hspace{0cm}\includegraphics[width=7cm]{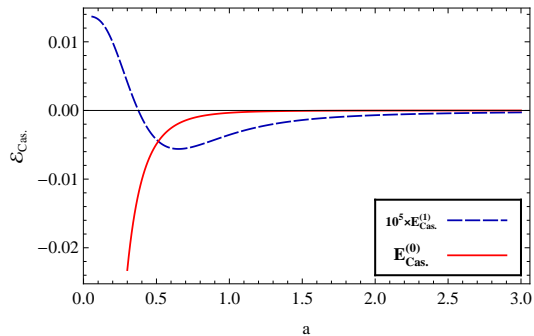}
  \caption{\label{figs.Casimir.}   The leading term for the Casimir energy and its first order radiative correction (multiplied by a factor of $10^5$) on two-dimensional spherical  surface with $S^2$ topology, are plotted as a function of the radius of sphere ($a$) for a massive scalar field  with $m = 1$ for $\lambda= 0.1$. Although, the sign of the values of correction term is changed, the zero order of Casimir energy is always negative.   }
\end{figure}

In this section, first order radiative correction of Casimir energy for the massive scalar field in $\phi^4$ theory on a spherical surface with $S^2$ topology is calculated. The counter-terms that we used in this calculation are the same as those in flat cases\,\cite{feynmanrule.1,feynmanrule.2,Toms.,L.S. Brown.1,L.S. Brown.2,hathrell.,bordag.linding.}. The renormalization procedure, the deduction of the counter-terms, and the final general form of the first order vacuum energy have been completely discussed in Refs.\,\cite{posision.dependent.counterterms.works.ours.1,posision.dependent.counterterms.works.ours.2}. Therefore, in this paper, we use only the important conclusions. Thus, we start with the perturbation expansion related to the two-point function as symbolically written as follows:
\begin{equation}\label{PerturbationEx.}
   \raisebox{-4mm}{\includegraphics[width=1.5cm]{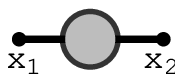}}=\raisebox{-2.7mm}{\includegraphics[width=1.4cm]{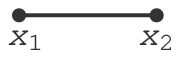}}
   +\raisebox{-2.5mm}{\includegraphics[width=1.2cm]{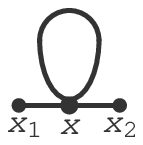}}+\raisebox{-3.5mm}{\includegraphics[width=1.5cm]{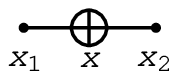}}\hspace{0cm}
\end{equation}
where $\raisebox{-3.mm}{\includegraphics[width=1.5cm]{16}}$ is the counter-term in the above perturbation expansion. By applying the appropriate renormalization conditions up to the first order of $\lambda$, the general expression for counter-terms is obtained as:
\begin{equation}\label{Counter-terms.}
   \delta_z=0,\hspace{2cm}\delta_m(x)=\frac{-i}{2}\raisebox{0.2mm}{\includegraphics[width=1cm]{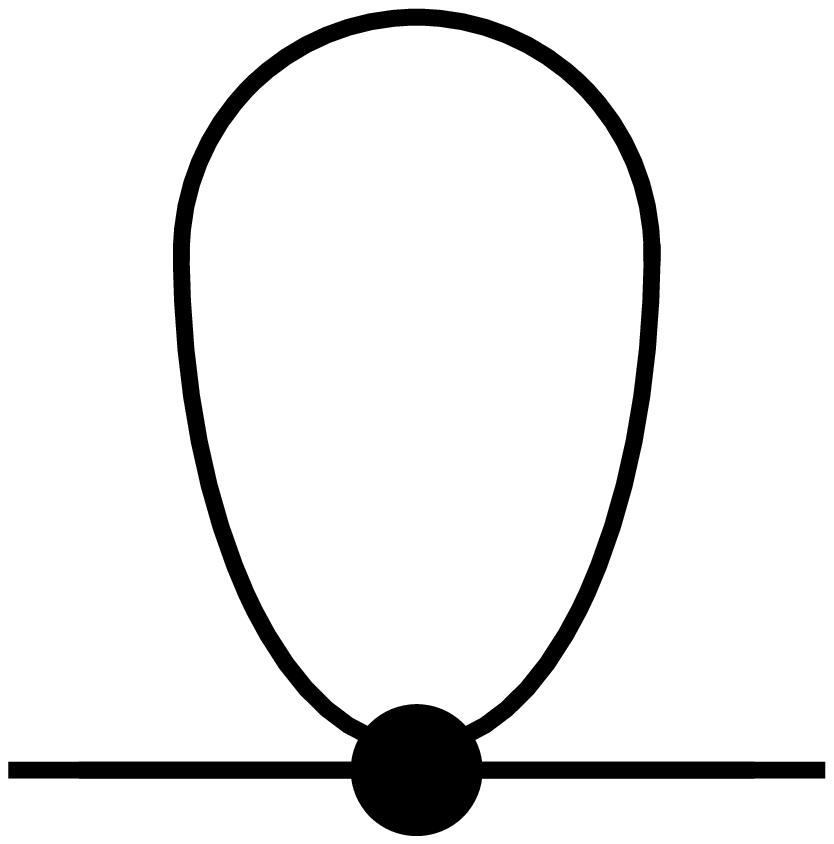}}=\frac{-\lambda}{2}G(x,x),\hspace{2cm}\delta_\lambda=0.
\end{equation}
In addition, the general expression for the first order vacuum energy is:
\begin{equation}\label{VacuumEn.damble.}
  E^{(1)}_A=i\int d^2\mathbf{x}\bigg(\frac{1}{8} \raisebox{-7mm}{\includegraphics[width=0.5cm]{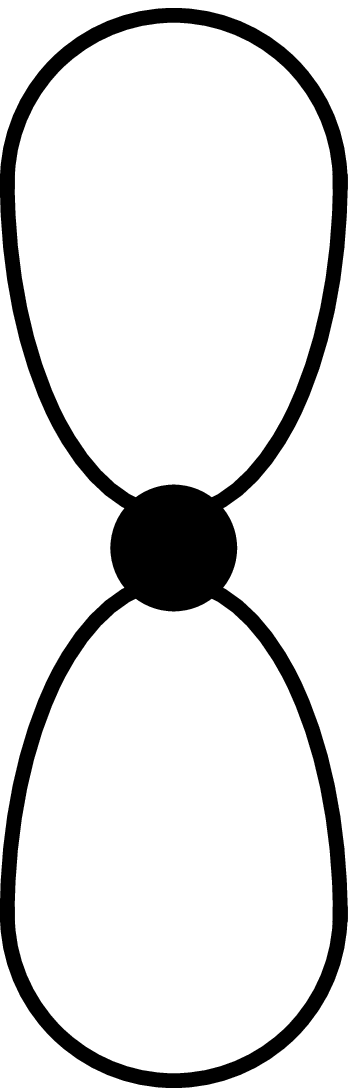}}+\frac{1}{2}\raisebox{-1mm}{\includegraphics[width=0.5cm]{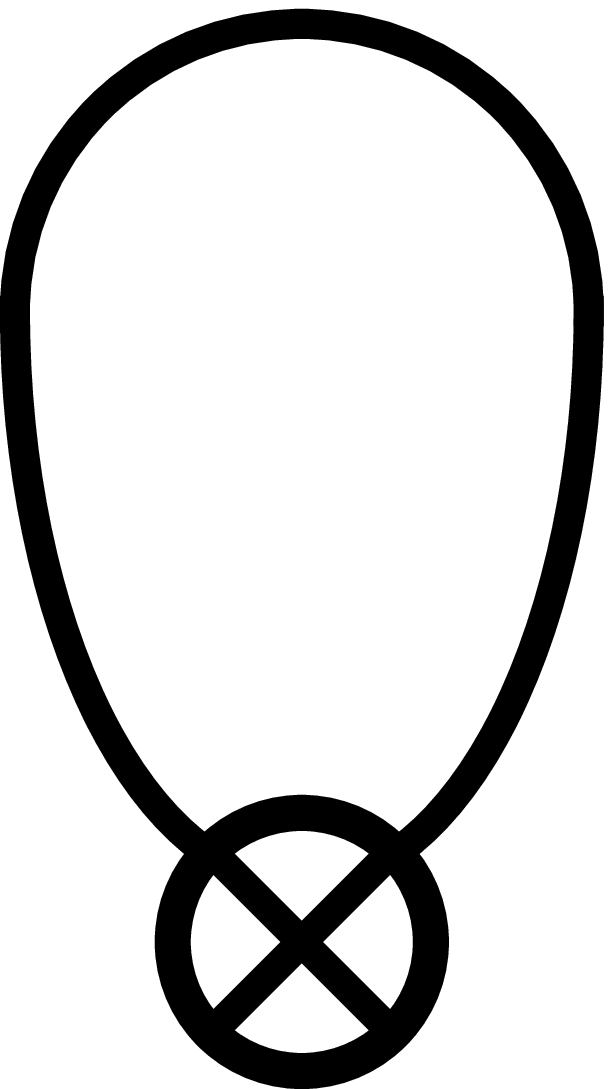}}+...\bigg)=i \int_{S}d^2\mathbf{x}\bigg(\frac{-i\lambda}{8}G^2(x,x)-\frac{-i}{2}\delta_m(x)G(x,x)\bigg)
\end{equation}
where $G(x,x)$ is the Green's function and the superscript $(1)$ denotes the first order of this energy. By substituting $\delta_m (x)$ from Eq.\,\eqref{Counter-terms.} for Eq.\,\eqref{VacuumEn.damble.} the total vacuum energy of configuration A is obtained. Therefore, we have:
\begin{equation}\label{VacuumEn.firstorder.}
  E^{(1)}_A=\frac{-\lambda}{8} \int_{S} G^2 (x,x)d^2\mathbf{x}
\end{equation}
According to the lagrangian stated in Eq.\,\eqref{lagrangy.} and using the usual computation, the final expression of the Green's function for real scalar field with the periodic boundary condition on the spherical surface with a radius $a$ after Wick rotation can be written as follows:
\begin{equation}\label{Green.function}
  G(x,x')=\int_{-\infty}^{\infty}\frac{d\omega}{2\pi}\sum_{\ell=0}^{\infty}\sum_{M=-\ell}^{\ell}\frac{e^{-\omega(t-t')}Y^M_\ell(\theta,\varphi)Y^{M*}_\ell(\theta',\varphi')}{\omega^2+\omega_\ell^2}
\end{equation}
where $\omega_\ell=\left[\frac{1}{a^2}\left(\ell+\frac{1}{2}\right)^2+m^2\right]^{\frac{1}{2}}$ and $Y_\ell^M(\theta,\varphi)$ is the spherical harmonic function. Furthermore, $x=(t,\theta,\varphi)$ shows the space-time coordinates on the surface. Now, after calculating the summation on $M$ and integral over $\omega$ in Eq.\,\eqref{Green.function}, the obtained result is substituted for Eq.\,\eqref{VacuumEn.firstorder.}. Therefore, for Eq.\,\eqref{VacuumEn.firstorder.}, we have:
\begin{equation}\label{VacuumEn.aftersumM.}
  E^{(1)}_A=\frac{-\lambda}{8}\int_{0}^{2\pi}\int_{0}^{\pi}\bigg(\sum_{\ell=0}^{\infty}\frac{2\ell+1}{8\pi\omega_\ell}\bigg)^2
  a^2\sin\theta d\theta d\varphi
\end{equation}
Now by computing the integral in Eq.\eqref{VacuumEn.aftersumM.} over the two coordinates $\theta$ and $\varphi$ on sphere, the final expression for the vacuum energy of configuration A is calculated. Thus, we have:
\begin{equation}\label{VacuumEn.after.int.space}
   E^{(1)}_A=\frac{-\lambda a^2}{32\pi}\bigg(\sum_{\ell=0}^{\infty}\frac{\ell+\frac{1}{2}}{\omega_\ell}\bigg)^2
\end{equation}
Since the above summation expression is divergent, a regularization technique is required. Thus, similar to what happened for Eq.\,\eqref{vacuum.energy}, the APSF for half integer parameters should be used\,\cite{Generalized.Abel.Plana.Saharian}. Therefore, the vacuum energy density $\mathcal{E}_A^{(1)}$ from Eq.\,\eqref{VacuumEn.after.int.space} is transformed into the following form:
\begin{equation}\label{VacuumEn.after.APSF}
  \mathcal{E}^{(1)}_A=\frac{-\lambda}{128\pi^2}\bigg(a\int_{0}^{\infty}\frac{t}{\sqrt{\mu^2+t^2}}dt+B(a)\bigg)^2
\end{equation}
where $\mu=ma$ and $\dd B(x)=2\mu a\int_{0}^{1}\frac{tdt}{\sqrt{1-t^2}(e^{2\pi \mu t}+1)}$ is the \emph{Branch-cut} term of APSF and it usually has a finite value. However, the first term in the parentheses in Eq.\,\eqref{VacuumEn.after.APSF} is divergent. In order to regularize and remove the divergences due to this term, the BSS is reused. Thus, according to the Casimir energy definition stated in Eq.\,\eqref{ECasDefinition.}, two similar configurations (as shown in Fig.(\ref{figs.sphere})) are introduced and the vacuum energies of these configurations in first order of $\lambda$ are subtracted from each other in proper limits.  Now, by substituting $a$ for $b$ in Eq.\,\eqref{VacuumEn.after.APSF} vacuum energy density $\mathcal{E}_B^{(1)}$ for configuration B is obtained. In the following, by using Eq.\,\eqref{ECasDefinition.}, we have:
\begin{equation}\label{first.order.Diff}
  \mathcal{E}^{(1)}_A-\mathcal{E}^{(1)}_B=\frac{-\lambda}{128\pi^2}\bigg\{a\int_{0}^{\infty}\frac{t}{\sqrt{\mu^2+t^2}}dt+B(a)\bigg\}^2-\{a\rightarrow b\}
\end{equation}
To remove infinities due to the first integral in both brackets of the above expression, the upper limits of integrations should be replaced with two multiple cutoffs (\emph{e.g.} $\Lambda_a$ and  $\Lambda_b$).  Similar to what occurs in Eq.\,\eqref{Vacuum.En.Diff.}, enough degrees of freedom to select appropriate cutoffs is accessible. This freedom in the selection of cutoffs allows the integration results to be regularized properly. Therefore, all infinities in Eq.\,\eqref{first.order.Diff} will be canceled via subtraction process and only the remaining terms are the Branch-cut terms and we have:
\begin{equation}\label{first.order.remaining.Branch.cut}
  \mathcal{E}^{(1)}_A-\mathcal{E}^{(1)}_B=\frac{-\lambda}{128\pi^2}\bigg[B(a)^2-B(b)^2\bigg]
\end{equation}	
Unfortunately, finding a closed form answer for the integral in Branch-cut term $B(x)$ is extremely difficult. Therefore, by expanding the exponential part of denominator in the integrand and computing the integral, the following result is obtained:
\begin{equation}\label{Branch-cut.firstorder}
   B(x)=\sum_{j=1}^{\infty}(-1)^{j+1}\pi m x^2[-I_1(2\pi m x j)+L_{-1}(2\pi m x j)]
\end{equation}
At the last step, the limit stated in Eq.\eqref{ECasDefinition.} is calculated and the expression for the first order radiative correction to the Casimir energy density becomes:
\begin{equation}\label{First.Order.Casimir.Expression.}
  \mathcal{E}^{(1)}_{\text{Cas.}}=\frac{-\lambda}{128\pi^2}\bigg[B(a)^2-\Big(\frac{1}{24m}\Big)^2\bigg]
\end{equation}
To calculate the Casimir energy for the massless case, we go back to the expression given in Eq.\,\eqref{first.order.remaining.Branch.cut} and we use $m$ as a regulator in this limit. Fortunately, there is no essential singularity and the value of the first order radiative correction to the Casimir energy in massless case is zero. Analogous to this way, by starting from Eq.\,\eqref{Leading.order.CasEn.}, it can be shown that the leading order of the Casimir energy in the massless limit is also zero. In Fig.\,(\ref{figs.Casimir.}), all the values for the zero order and the first order radiative correction to the Casimir energy for a massive scalar field ($m = 1$) as a function of the radius of sphere were plotted. We should mention that the correction terms are approximately $10^5$ times smaller than their zero order counterparts. This plot also shows, the Casimir energy values for both orders of correction diminishes when the scale of radius of sphere goes infinity. This behavior of the Casimir energy is fortunately consistent with the expected physical grounds.

\section{Conclusions}
\label{sec:conclusion}
In this paper, the leading order and first order of radiative correction to the Casimir energy for massive scalar field in  $\phi^4$ theory on a curved manifold with $S^2$ topology were calculated. The major point distinguishing this paper from others is performing a different renormalization program in the calculation process. We maintain that the renormalization program should completely and self-consistently take account of the boundary conditions or any possible nontrivial background breaking the translational invariance of the system. In other words, for the renormalization of the physical parameters in the presence of the boundary conditions or nontrivial backgrounds, the counter-term related to the free theory should not be imported in problems. To be more specific, the counter-terms should be automatically obtained from the theory of perturbation expansions. This work guarantees that the obtained counter-terms reflect all information about the boundary conditions imposed or possible non-trivial background. This program for renormalization has been performed previously in multiple studies\,\cite{posision.dependent.counterterms.works.ours.1,posision.dependent.counterterms.works.ours.2} and its various aspects are also explained before. In this paper, for the first time, we used the mentioned renormalization program on a spherical surface as a curved manifold. In addition to the renormalization program, to regularize and remove divergences in calculation, the BSS as a supplementary method was performed. In this scheme, two similar configurations were introduced and to compute the Casimir energy, the values of vacuum energy for these two configurations are subtracted from each other. Using this procedure removes all the ambiguities associated with the appearance of the infinities. In this paper, the BSS was first used in the computation of the leading order of the Casimir energy. Then, using the aforementioned renormalization program along with the BSS, the first order radiative correction to the Casimir energy was calculated. The obtained results for both zeroth and first orders of radiative corrections satisfy the expected physical grounds well and for the massless case, the Casimir energy for both orders is zero.
\acknowledgments
The Author would like to thank the research office
of Semnan Branch, Islamic Azad University for the financial support.

\end{document}